# How are exclusively data journals indexed in major scholarly databases? An examination of the Web of Science, Scopus, Dimensions, and OpenAlex

**Authors:**
Chenyue Jiao (School of Information Sciences, University of Illinois Urbana-Champaign, 501 E. Daniel St., Champaign, IL, 61820)
Kai Li (University of Tennessee, Knoxville, 1345 Circle Park Drive 451 Communications Building, Knoxville, TN 37996)
Zhichao Fang (Renmin University of China, School of Information Resource Management, Renmin University of China, School of Information Resource Management, Beijing, China 100872; Leiden University, Centre for Science and Technology Studies, Kolffpad 1, 2333 BN Leiden, the Netherlands)

Correspondence: Kai Li kli16@utk.edu

**Abstract:**
As part of the data-driven paradigm and open science movement, the data paper is becoming a popular way for researchers to publish their research data, based on academic norms that cross knowledge domains. Data journals have also been created to host this new academic genre. The growing number of data papers and journals has made them an important large-scale data source for understanding how research data is published and reused in our research system. One barrier to this research agenda is a lack of knowledge as to how data journals and their publications are indexed in the scholarly databases used for quantitative analysis. To address this gap, this study examines how a list of 18 exclusively data journals (i.e., journals that primarily accept data papers) are indexed in four popular scholarly databases: the Web of Science, Scopus, Dimensions, and OpenAlex. We investigate how comprehensively these databases cover the selected data journals and, in particular, how they present the document type information of data papers. We find that the coverage of data papers, as well as their document type information, is highly inconsistent across databases, which creates major challenges for future efforts to study them quantitatively. As a result, we argue that efforts should be made by data journals and databases to improve the quality of metadata for this emerging genre.

# 1 Introduction

Research data has become one of the most important objects in the research system during the past decade. Researchers across knowledge domains are relying on larger quantities of data to understand their research topics, which has brought significant changes to how our research system works and how research is conducted (Borgman, 2012; Hey et al., 2009). In particular, it is commonly agreed that the increasing amount of research data has raised distinct new requirements for data collection, processing, publishing, and sharing (Chen et al., 2013), which cannot be sufficiently fulfilled without support from new infrastructure (Chawinga & Zinn, 2019). One recent development in this area is the 2016 proposal of the FAIR principles as guidelines for various stakeholders in the e-Science domain to enhance the findability and usability of data objects (Wilkinson et al., 2016). The 15 principles form a clear and actionable framework for the development of data-related initiatives and have been embraced by many parties in the research community.

Another significant recent development concerning research data is the academic genre of the data paper, which gradually took shape in the early 2010s as a "scholarly publication of a searchable metadata document describing a particular online accessible dataset, or a group of datasets, published in accordance to the standard academic practices" (Chavan & Penev, 2011, p. 3). It serves as a descriptor and citable proxy of data objects in the bibliographic universe, so that research data can be more findable, citable, and reusable under the current research infrastructure (Gorgolewski et al., 2013; Li et al., 2020; Li & Jiao, 2021), goals that are consistent with the FAIR principles (Groth et al., 2020; Schöpfel et al., 2020). Moreover, data papers are making it easier for research data to be peer-reviewed, a significant prerequisite for the integration of data objects into the research system (Costello et al., 2013; Mayernik et al., 2015). Over time, more journals have begun accepting data papers; these periodicals are termed data journals (Candela et al., 2015).

As data papers are becoming a popular way for researchers to publish their research data in many disciplines (Candela et al., 2015; Griffiths, 2009), this new genre has become an important data source for investigating how data is used by scientists. This echoes increasing interest in research data from the field of quantitative science studies (Cousijn et al., 2019; Silvello, 2018). Numerous studies have been conducted using quantitative methods and large-scale datasets to understand the relationship between research data and scientific studies and outputs, such as how data objects are cited and/or mentioned in scientific publications (Färber et al., 2021; Lafia et al., 2021; Zhao et al., 2018) and the disciplines behind the datasets (Fan et al., 2023). The majority of existing research uses citations to data repositories, such as DataCite (Robinson-Garcia et al., 2017) and the Inter-university Consortium for Political and Social Research (ICPSR) data repository (Fan et al., 2023; Lafia et al., 2022) as well as Clarivate's Data Citation Index (Park & Wolfram, 2019; Robinson-García et al., 2016), which is also primarily based on data repositories (Force & Robinson, 2014). However, despite the growing importance of data papers, very few studies in this line of research have analyzed them directly, with a few exceptions based on small numbers of individual data journals (Li et al., 2020; Li & Jiao, 2021; McGillivray et al., 2022).

The absence of data papers from large-scale empirical studies represents a major gap in the existing research infrastructure for effectively tracing data papers. Efforts have been made to identify data journals (Candela et al., 2015; Walters, 2020), but to our knowledge, no research has been conducted to understand how these journals and their publications are indexed in scholarly databases, such as the Web of Science and Scopus, which are frequently used as the direct data source in quantitative science studies. This gap makes it harder for researchers to easily extract a large body of data papers from the scholarly databases and analyze them, especially using quantitative methods.

To bridge this important gap, this research aims to examine the coverage of data journals and papers indexed in major scholarly databases used in quantitative science studies, including the Web of Science (WoS), Scopus, Dimensions, and OpenAlex. We selected a list of *exclusively data journals* (i.e., journals that only or primarily accept data papers) from lists of data journals that have been compiled by other researchers. Using this list, we evaluated how data papers in these journals are indexed in the above databases, particularly from the perspectives of document types used to describe the publications and changes in coverage over time. More specific research questions include:

**RQ1: Which exclusively data journals are indexed in major scholarly databases?** Using various lists of data journals (discussed in Methods), we compile a list of exclusively data journals based on our operationalization of this concept and quantitatively examine their presence in the databases listed above. This will serve as the basis for future quantitative studies on the genre of data papers.

**RQ2: How are they indexed over time?** Building upon the survey of data journals in RQ1, we further examine how different databases index exclusively data journals over time, to understand the coverage of this genre from a more granular and dynamic perspective.

**RQ3: Are they indexed accurately in terms of document type?** The last question aims to offer a survey of the extent to which data papers in the journals are labelled as data papers in the selected databases. Correct labelling is the first step for data papers to be distinguished from other types of publications (especially research articles) in a database and analyzed separately. Answers to this research question will lead to a better understanding of the gaps in the current infrastructure for data papers and facilitate more meaningful support of data publication in the future.

## 2 Literature Review

### 2.1 Data publication and citation

The role of data in scientific research has been accentuated by the data-driven research paradigm from the beginning of the 21$^{st}$ century, where researchers must make use of increasing amounts of research data to create new knowledge (Hey et al., 2009). Researchers have proposed that data has been, or at least should be, treated as a first-class research object (Callaghan et al., 2012; Starr et al., 2015), one that on a similar footing with research articles. This requirement reflects how research data should be handled in research projects, especially along the lifecycle of research data sharing, peer review, citation, and reuse.

Publishing data, or sharing data in a way that complies with academic norms, has already become commonplace, despite the different interpretations of this concept (Kratz & Strasser, 2014). A large number of data repositories have been established to make data publicly available (Kim & Burns, 2016; Kowalczyk & Shankar, 2011). More recently, researchers have also embraced the practice of publishing a paper to describe the dataset (Jiao & Darch, 2020). Serving as a metadata document for research data and a vehicle for the dataset to be peer reviewed, the data paper helps to incentivize researchers to publish their data and integrate research data into the academic system (Chavan & Penev, 2011; Li & Jiao, 2021). Moreover, by taking the form of an academic document, data papers are more easily cited and rewarded in the research system as it currently exists (Huang & Jeng, 2022).

The popularity of this new academic genre is seen in the number of data journals created to publish data papers, whether exclusively or mixed with other types of documents. Candela and colleagues' survey (2015) identified more than 100 data journals publishing data papers by the year 2015. Yet the landscape is fluid: new journals are being established every year, and many of these data journals have ceased to publish as well (Schöpfel et al., 2020; Stuart, 2017).

The increasing availability of data repositories and data papers has opened new possibilities for quantitative researchers who seek to investigate the relationship between research data and the sciences from a grander perspective (Peters et al., 2016). However, the majority of existing research is based on various data repositories, such as DataCite (Robinson-Garcia et al., 2017), the Inter-university Consortium for Political and Social Research (ICPSR) data repository (Fan et al., 2023; Lafia et al., 2022) and Clarivate's Data Citation Index (Park & Wolfram, 2019; Robinson-García et al., 2016), which is also primarily based on data repositories (Force & Robinson, 2014). In contrast, very few studies are focused on data papers and journals, and these exceptions mostly use individual data journals to draw smaller-scale conclusions about the publication and reuse of data objects (Li & Jiao, 2021; McGillivray et al., 2022; Thelwall, 2020). This limitation is primarily due to the lack of evidence about how data journals and papers are indexed in scholarly databases, which is the direct motivation of the present study.

## 2.2 Coverage of scholarly databases

Scholarly databases provide important infrastructure for quantitative studies of science, as they offer easily-to-access citation data and publication metadata to be used in large-scale analyses. Serving as important data sources, they can strongly define the scope and quality of data used by such studies (Mingers & Leydesdorff, 2015). In the current landscape of scholarly databases, the Web of Science (Li et al., 2018) and Scopus (Burnham, 2006) are the two earliest and most established examples, created in the late 1990s and early 2000s, respectively. Both of these databases use more selective criteria to determine the list of journals to be included. Google Scholar, created in 2004, represents a distinct approach from the foregoing and has received both positive and negative evaluations from researchers (Jacsó, 2005). Since the 2010s, Microsoft Academic Graph (MAG) and Dimensions have also joined the market and become heavily used (Singh et al., 2021). After MAG was discontinued in 2021, a new knowledge-graph-based database, OpenAlex, was developed as a replacement (Priem et al., 2022) and has shown strong potential for future quantitative studies of science (Scheidsteger & Haunschild, 2023).

These databases differ from each other in many respects—most notably the scope of publications included, which heavily shapes the results that can be derived from them. As a result, numerous empirical studies have been conducted to understand the coverage of these databases, which forms a major line of research in quantitative studies of science. Such studies operate mainly through comparison, on the level of journals (Gavel & Iselid, 2008; Mongeon & Paul-Hus, 2016; Singh et al., 2021) and of articles and their citations (Archambault, 2009; De Groote & Raszewski, 2012; Harzing & Alakangas, 2016; Martín-Martín et al., 2018, 2021), with some research investigating the growth of resources over time (Larsen & von Ins, 2010; Michels & Schmoch, 2012). Moreover, comparisons have been made of specific types of resources in the database, such as publications in non-English languages (Moed et al., 2018; van Leeuwen et al., 2001; Vera-Baceta et al., 2019) as well as document types other than the research article (Meho & Sugimoto, 2009; Michels & Fu, 2014; Visser et al., 2021). Most of this research has confirmed that newer databases are more comprehensive than the two established databases from most perspectives, but especially in terms of non-traditional resources, such as conference and non-English publications.

Another aspect of this comparison is the quality of metadata, especially the accuracy of document type assignment, which is important for the selection of publications for quantitative analysis using these general-purpose databases (Donner, 2017), but particularly important for the identification of data papers from these databases. Studies have found that the Web of Science and Scopus tend to have more granular and accurate document type classifications (Singh et al., 2021; Visser et al., 2021). However, marginalized document types, such as reviews, letters, and notes, do not have 100% accuracy in any database (Harzing, 2013). Given the importance of document type to queries for data papers, this research is designed to understand this issue as it applies to the case of data papers.

## 3 Methods

### 3.1 Identifying exclusively data journals

Data journals, as a new venue for data sharing and publishing, have gained increasing attention from scholars. There are many resources that provide lists of data journals. In this study, we resorted to the following resources to compile a list of exclusively data journals: (1) Candela and colleagues' survey (Candela et al., 2015), (2) an updated journal list by Walters (2020), (3) a list of data journals compiled by Kindling and Strecker (2022), (4) data journal lists created by academic libraries and other parties indexed by Google (e.g., the list of data journals created by the University of Pittsburgh[1]), and (5) journals with "data" or "database" in the title included in the Journal Citation Reports or Scopus List of Journals.

From these sources, we further selected exclusively data journals based on the following criteria: (1) the journal primarily accepts data papers based on its statement of aims and scopes, operationalized as a greater than 50% share of data papers among all publications on the journal website, (2) the journal is active as of January 2023, and (3) the journal only publishes English-language articles. We manually examined all candidate journals against these criteria. For example, only about one-quarter of all publications in *Biodiversity Data Journal* are data papers, leading us to remove this periodical from the present study, despite the fact that it is mentioned as an important data journal in previous studies (Chavan & Penev, 2011; Li et al., 2020). We also excluded *Arxius de Miscellania Zoologica*, which publishes data papers in Catalan, English, and Spanish. *Genomics Data* was excluded because it published from 2013 to 2017; the journal is now part of *Data in Brief*.

---
[1] https://pitt.libguides.com/findingdata/datajournals

Finally, we selected the 18 journals shown in Table 1 as the analytical sample for this research.

**Table 1: Complete list of data journals in our sample**

| Data journal | Publisher | Initial year |
|---|---|---|
| Chemical Data Collections | Elsevier | 2016 |
| Data in Brief | Elsevier | 2014 |
| Earth System Science Data | Copernicus | 2009 |
| Freshwater Metadata Journal | Freshwater Information Platform | 2014 |
| Geoscience Data Journal | Wiley | 2014 |
| International Journal of Food Contamination | BMC | 2014 |
| IUCrData | International Union of Crystallography | 2016 |
| Journal of Chemical and Engineering Data | American Chemical Society | 1959 |
| Journal of Open Archaeology Data | Ubiquity Press | 2012 |
| Journal of Open Humanities Data | Ubiquity Press | 2015 |
| Journal of Open Psychology Data | Ubiquity Press | 2013 |
| Journal of Physical and Chemical Reference Data | AIP Publishing | 1972 |
| Nuclear Data Sheets | Elsevier | 1971 |
| Open Data Journal for Agricultural Research | —[2] | 2015 |
| Open Health Data | Ubiquity Press | 2013 |
| Open Journal of Bioresources | Ubiquity Press | 2014 |
| Research Data Journal for the Humanities and Social Sciences | Brill | 2016 |
| Scientific Data | Nature | 2014 |

**3.2 Collecting data papers from major databases**

In this study, we strove to answer our research questions using the Web of Science (WoS), Scopus, Dimensions, and OpenAlex, as these are among the most commonly used large-scale bibliographic data sources in quantitative science studies.

---

[2] This independently published journal is sponsored by Wageningen University and Research Centre Library.

For each database, we collected the metadata information of all publications from each journal in our final list that was indexed in the database. We used the online portal of the Web of Science and Scopus for data collection. As for Dimensions and OpenAlex, we retrieved the data from the in-house Dimensions database (version: June 2022) and OpenAlex database (version: October 2022) hosted at the Centre for Science and Technology Studies (CWTS) of Leiden University, respectively. We only considered papers published by the end of 2021.

### 3.3 Comparing document types of our sample

The classification of document types varies by database. In Dimensions and OpenAlex, all publications from indexed data journals are classified as *Article*, whereas various document types appear in WoS and Scopus. Therefore, we only compared the document types between the latter two databases. Table 2 shows how the two main document types of interest, *Article* and *Data paper*, are defined by these two databases, as quoted from their documentation. Based on their definitions and their presentation in the data, we classified publications into these two types. Our classification also includes other document types, such as *Correction* and *Editorial material*, which are categorized as *Other* in this research. We note that, based on our examination, the WoS retrospectively assigned *Data paper* to articles published before 2016, when this type was introduced. However, it was unclear when the *Data paper* tag was introduced into Scopus. As a result, our analysis of the document type must be based on the data collected at this time.

**Table 2: Document type policies in WoS and Scopus**

| Database | Document type | Description |
|---|---|---|
| WoS[3] | Article | Reports of research on new and original works that are considered citable. Includes research papers, brief communications, technical notes, chronologies, full papers, and case reports (presented like full papers) that were published in a journal and/or presented at a symposium or conference. Articles usually include author abstract, graphs, tables, and lists of cited references. |
| WoS[3] | Data paper | A scholarly publication describing a particular dataset or collection of datasets and usually published in the form of a peer-reviewed article in a scholarly journal. The main purpose of a data paper is to provide facts about the data (metadata, such as data collection, access, features etc.) rather than analysis and research in support of the data, as found in a conventional research article. **A Data Paper will have a dual document type: Article; Data Paper**. **Prior to 2016, a Data Paper was processed as an Article only**. |
| Scopus[4] | Article | Original research or opinion. Articles in peer-reviewed journals are usually several pages in length, most often subdivided into sections: abstract, introduction, materials & methods, results, conclusions, discussion, and references. However, case reports, technical and research notes and short communications are also considered to be |

---

[3] https://webofscience.help.clarivate.com/en-us/Content/document-types.html
[4] https://www.elsevier.com/__data/assets/pdf_file/0007/69451/ScopusContentCoverageGuideWEB.pdf

| | | |
|---|---|---|
| | | articles and may be as short as one page in length. Articles in trade journals are typically shorter than in peer-reviewed journals, and may also be as brief as one page in length. |
| | Data paper | Searchable metadata documents describing an online accessible dataset, or group of datasets. The intent of a data paper is to offer descriptive information on the related dataset(s) focusing on data collection, distinguishing features, access, and potential reuse rather than information on data processing and analysis. |

To examine the accuracy of document types in these databases, we collected the papers' classification on the journal website and compared it with the document types assigned by the two databases. We focused on one journal in our list, *Scientific Data*, as a case study for two reasons: first, *Scientific Data* is the most influential data journal, especially in terms of impact factor; second, publications in this journal are searchable by article type on the website[5] so that it is easy to collect the classifications of each publication.

We collected 1,913 publications published by the end of 2021 in *Scientific Data* for this case study. The document types defined by the journal and the count of publications in each type are presented in Table 3. According to the definitions, we counted *Data descriptor* as *Data paper*, *Article* as *Article*, and all remaining categories as *Other* in our analysis. We then compared the count of publications in these categories from the journal website, WoS, and Scopus to examine the extent to which data journal publications are classified correctly in databases.

**Table 3: Document types in *Scientific Data***

| Document type used by *Scientific Data* | Publications | Category definition[6] |
|---|---|---|
| Data descriptor | 1636 | Detailed descriptions of research datasets, which focus on helping others reuse data, rather than testing hypotheses or presenting new interpretations. |
| Article | 69 | Reports on new policies, repositories, standards, ontologies, workflows, or any topic relating to the mechanics of data sharing. |
| Author correction | 53 | / |
| Analysis | 44 | A new analysis or meta-analysis of existing data, which highlights examples of data reuse or new findings. |
| Comment | 41 | Short commentaries or opinions on research data policy, workflows or infrastructure that don't need to report a specific technology or finding. |
| Corrigendum | 22 | / |
| Publisher correction | 17 | / |
| Editorial | 14 | / |

---

[5] https://www.nature.com/sdata/articles
[6] https://www.nature.com/sdata/publish/submission-guidelines#sec-1

| | | | |
|---|---|---|---|
| Erratum | 10 | | / |
| Addendum | 6 | | / |
| Retraction | 1 | | / |
| Total | 1913 | | / |

## 4 Findings

### 4.1 How are data journals and publications indexed?

The numbers of data journals and the year in which they were indexed in the database vary significantly among these four popular scholarly databases (see Table 4). Only eight data journals are indexed in WoS and 11 in Scopus, but Dimensions and OpenAlex have full coverage of the journal list. In terms of the indexed year, even though most of the journals are indexed in these databases in very similar manners, there are some notable differences, the majority of which are due to the fact that WoS is the most selective database among these four. Another notable observation is that the indexed years of *Journal of Chemical and Engineering Data* in three databases are prior to its established year. This is because a few publications were indexed by their first published dates instead of their formal published dates. Additionally, the fact that the WoS has later indexed years than most of the other databases is consistent with the fact that it has the most selective criteria for journals among the most popular databases (Norris & Oppenheim, 2007).

**Table 4: Indexing of data journals in the four databases**

| Data journal | Year established | First year of indexing | | | |
|---|---|---|---|---|---|
| | | WoS | Scopus | Dimensions | OpenAlex |
| Chemical Data Collections | 2016 | | 2016 | 2016 | 2016 |
| Data in Brief | 2014 | 2018 | 2014 | 2014 | 2014 |
| Earth System Science Data | 2009 | 2012 | 2009 | 2009 | 2009 |
| Freshwater Metadata Journal | 2014 | | | 2014 | 2014 |
| Geoscience Data Journal | 2014 | 2014 | 2015 | 2012 | 2012 |
| International Journal of Food Contamination | 2014 | | 2014 | 2014 | 2014 |
| IUCrData[7] | 2016 | | | 2016 | 2016 |
| Journal of Chemical and Engineering Data | 1959 | 1965 | 1956 | 1956 | 1955 |
| Journal of Open Archaeology Data | 2012 | 2018 | | 2012 | 2012 |
| Journal of Open Humanities Data | 2015 | | | 2015 | 2015 |
| Journal of Open Psychology Data | 2013 | | | 2013 | 2013 |
| Journal of Physical and Chemical Reference Data | 1972 | 1977 | 1972 | 1972 | 1972 |
| Nuclear Data Sheets | 1971 | 2003 | 1971 | 1971 | 1971 |

---

[7] *IUCrData* and *Journal of Open Humanities Data* were indexed in Scopus from 2022, which is not covered by our publication window.

| | | | | | |
|---|---|---|---|---|---|
| Open Data Journal for Agricultural Research | 2015 | | | 2015 | 2015 |
| Open Health Data | 2013 | | | 2013 | 2014 |
| Open Journal of Bioresources | 2014 | | 2019 | 2014 | 2014 |
| Research Data Journal for the Humanities and Social Sciences | 2016 | | 2016 | 2016 | 2016 |
| Scientific Data | 2014 | 2014 | 2014 | 2014 | 2014 |

Table 5 presents the number of data journals established in three periods of time. We acknowledge that there may be other ways to classify the history of data journals; however, we selected the year 2014 because of its importance: multiple important data journals were established in this year, such as *Scientific Data* and *Data in Brief*. The earliest data journal is the *Journal of Chemical and Engineering Data* which was first published in 1959, followed by two others founded in the 1970s. We included these three data journals even though their inchoate publications may not be totally consistent with how data papers are defined today. A notable trend that can be observed from Table 5 is that most data journals in our list were established between 2014 and 2016, indicating that the data paper is a new and growing academic genre.

Table 5: Summary of the founding years of data journals

| Time period | Journals founded |
|---|---|
| Before 2000 | 3 |
| 2000-2013 | 4 |
| 2014-2016 | 11 |

Since the coverage and indexed year of data journals vary among databases, the number of publications also varies greatly. Table 6 illustrates the number of publications from each journal in the four databases. We found that OpenAlex has the most comprehensive coverage of publications, whereas the WoS has the fewest publications. For most journals, there is a small variance in the number of publications indexed in the databases, despite the identical publication window taken by these different databases. One notable issue we found in Table 6, compared to the data collected from *Scientific Data* (Table 3), is that three of the databases have more publications than the number of publications on the journal's website per se. We double-checked our data collection pipeline and found that the extra publications are primarily ascribable to indexing errors, where the same publication is assigned different IDs and/or titles. To highlight this quality issue, we decided not to remove the duplicated publications from our analytical sample.

Table 6: Publications from each journal indexed in the databases

| Data journal | WoS | Scopus | Dimensions | OpenAlex |
|---|---|---|---|---|
| Chemical Data Collections | / | 776 | 778 | 778 |
| Data in Brief | 5707 | 7567 | 7711 | 7823 |
| Earth System Science Data | 921 | 954 | 958 | 966 |
| Freshwater Metadata Journal | / | / | 51 | 52 |
| Geoscience Data Journal | 94 | 80 | 119 | 153 |
| International Journal of Food Contamination | / | 83 | 86 | 86 |
| IUCrData | / | / | 1536 | 1538 |
| Journal of Chemical and Engineering Data | 16910 | 18046 | 18088 | 18508 |
| Journal of Open Archaeology Data | 31 | / | 53 | 54 |
| Journal of Open Humanities Data | / | / | 51 | 51 |
| Journal of Open Psychology Data | / | / | 38 | 38 |
| Journal of Physical and Chemical Reference Data | 979 | 1039 | 1065 | 1090 |
| Nuclear Data Sheets | 802 | 1563 | 2607 | 2708 |
| Open Data Journal for Agricultural Research | / | / | 27 | 35 |
| Open Health Data | / | / | 26 | 22 |
| Open Journal of Bioresources | / | 27 | 50 | 50 |
| Research Data Journal for the Humanities and Social Sciences | / | 27 | 43 | 43 |
| Scientific Data | 1913 | 1943 | 1914 | 1943 |
| Total | 27357 | 32105 | 35201 | 35938 |

We further examined the numbers of journals and publications covered by each database over time. Figure 1 shows the trend on the journal level (Panel A) and the publication level (Panel B) respectively. We see a similar increasing trend for both journals and publications over time, especially from 2014 onwards. By the year of 2000, more than 15 exclusively data journals published more than 3,500 data papers every year, which shows the growth of this new academic genre. However, a notable difference between the databases can also be observed by comparing the two panels in Figure 1: even though the number of journals covered by Dimensions and OpenAlex is much larger than that of the other two, their indexed publications are similar in size. This is because most of the journals covered by WoS and Scopus are larger than those omitted; being more selective, these two databases included journals that are potentially more established and important. As a result, despite the fairly large difference in the number of data journals from these databases, we can still use the WoS and Scopus to retrieve a large enough and potentially more representative sample of data papers from these exclusively data journals.

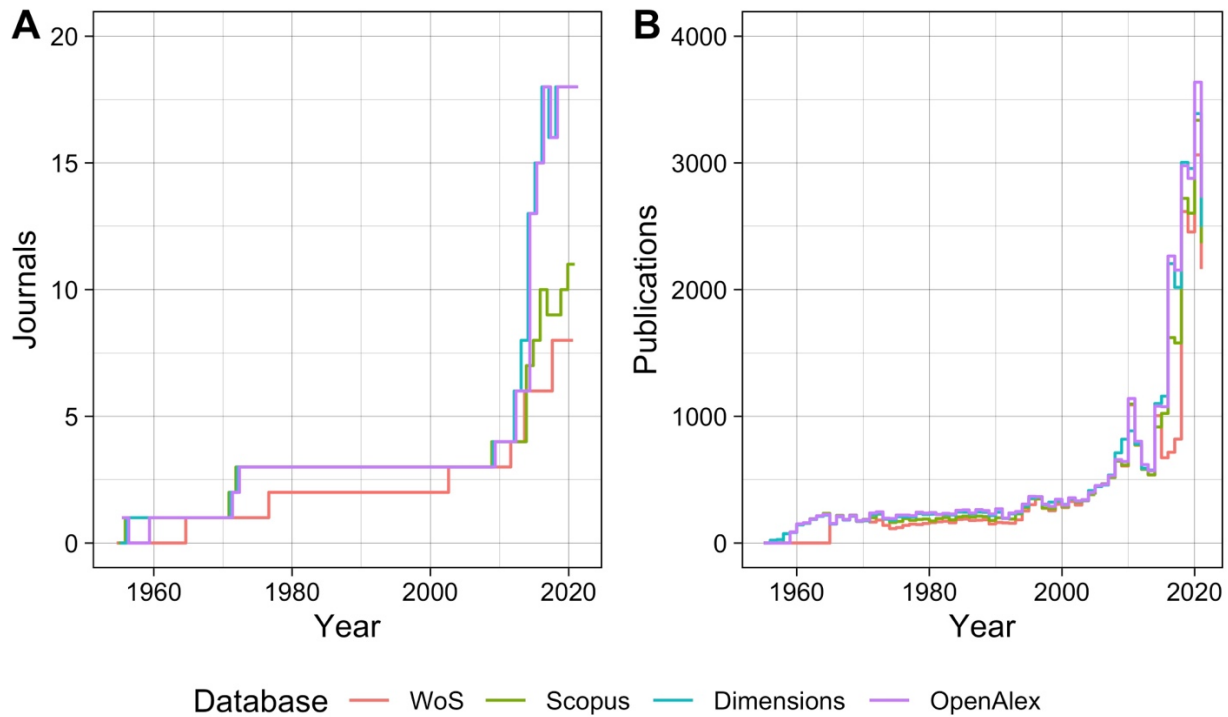

**Figure 1: Numbers of data journals (Panel A) and data papers (Panel B) indexed in the four databases over time**

### 4.2 Are publication types indexed consistently in the scholarly databases?

Each database has their own classification system of document types. Dimensions and OpenAlex assign *Article* to all data papers as they do for research articles, whereas WoS and Scopus have a specific category for data papers. Following the classification principles mentioned in Methods, Table 7 presents the share of all publications in each document type from the four databases. The distributions of publication in WoS and Scopus are similar to each other.

**Table 7: Share of all data papers across the three document types**

| Document type | Share in WoS | Share in Scopus | Share in Dimensions | Share in OpenAlex |
|---|---|---|---|---|
| Article | 66.90% | 66.34% | 100% | 100% |
| Data paper | 26.84% | 29.39% | / | / |
| Other | 6.25% | 4.27% | / | / |

We further evaluate the above trend from WoS and Scopus over time for four journals that are fully covered by both databases. Figure 2 shows that, despite the similar overall distributions in Table 7 between the two databases, there are vast differences in how document types are assigned in individual journals over time and between the two databases. This clearly shows that the assignment of the *Data paper* tag is far from consistent in any of these databases and cannot be reliably used to retrieve data papers in these two databases.

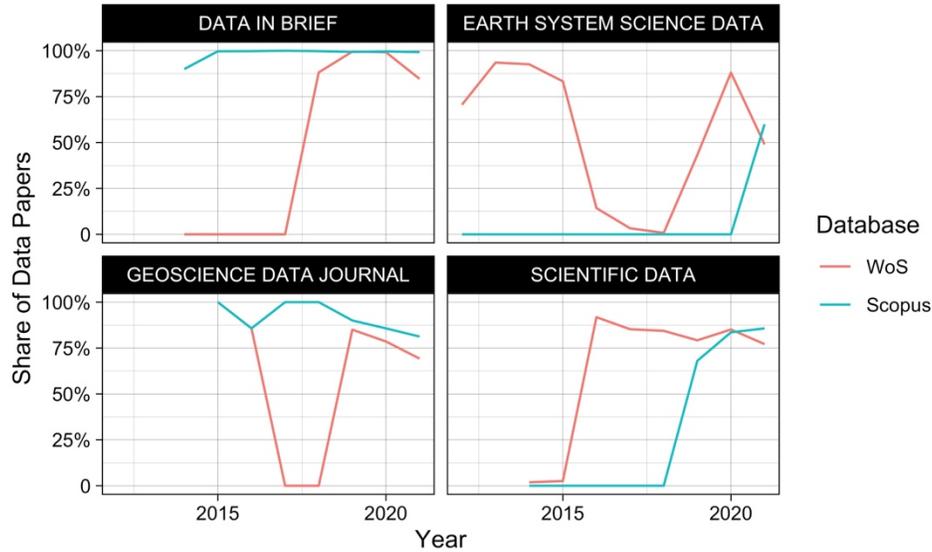

**Figure 2: Share of data papers in each journal in each database over time**

We further analyzed how publications from *Scientific Data* are indexed in WoS and Scopus, to understand the accuracy of document type assignment in a more granular manner. From the website of *Scientific Data*, there are 1,636 data papers, 69 articles, and 208 other publications, based on our classification. Table 8 shows how these publications are treated in the two databases. Even though both databases have many mislabeled articles, the WoS has a much higher accuracy (84.32%) than Scopus (59.27%).

**Table 8: Correspondence between WoS and Scopus document types and those on the *Scientific Data* website**

|  |  | **Document type from WoS** | | |
|---|---|---|---|---|
|  |  | Data paper | Article | Other |
| **Document type from journal** | Data paper | 1437 | 198 | 1 |
|  | Article | 23 | 46 | 0 |
|  | Other | 18 | 60 | 130 |

|  |  | **Document type from Scopus** | | |
|---|---|---|---|---|
|  |  | Data paper | Article | Other |
| **Document type from journal** | Data paper | 903 | 719 | 14 |
|  | Article | 0 | 69 | 0 |
|  | Other | 0 | 45 | 162 |

# 5 Discussion

In this work, we analyzed how exclusively data journals are indexed in major scholarly databases, as a first step towards establishing a comprehensive sample of data papers for future quantitative analyses. More specifically, we compiled a list of 18 exclusively data journals using existing efforts and analyzed how these journals and their publications are indexed and labelled by four such databases. Our results show significant inconsistencies in the indexing and labeling of data journals and papers by the popular databases, a major gap to be addressed by future efforts to improve the infrastructure that supports data publication and citation.

On the journal level, our results show that the two newer databases, Dimensions and OpenAlex, enjoy a strong advantage over the two more traditional databases, WoS and Scopus. The former two databases cover all of the exclusively data journals, whereas Scopus and WoS only cover 11 and 8 journals, respectively. Our results echo findings from past research that new databases in the market, such as Microsoft Academic Graph (the predecessor of OpenAlex) and Dimensions, are generally more comprehensive in terms of the research outputs indexed (Singh et al., 2021; Visser et al., 2021). This trend is especially applicable to data journals because many of the exclusively data journals are relatively new and do not have many publications and citations, which makes these journals much less likely to be indexed in more established and selective databases.

Despite the large difference in the number of journals covered by these databases, we also find that the numbers of articles covered by the databases are much more similar to each other. Scopus and WoS covers about 90% and 75% of all articles in OpenAlex, respectively. This is because most of the data journals indexed in Scopus and WoS are also those with higher impact and more publications. This also shows that we will be able to collect a "good enough" and potentially more representative sample of data papers by simply using Scopus and, to a lesser extent, WoS.

Beyond the data journals and papers indexed in the databases, we also examined the document type tag used for data papers in these databases, as this is the metadata element that will need to be used to retrieve data articles. Among the four databases we examined, all data papers are counted as regular research articles in Dimensions and OpenAlex, making it very challenging for researchers to acquire a full sample of data papers from them, despite their more comprehensive coverage of data papers. This is consistent with existing empirical evidence that the document type tag suffers from quality issues in most scholarly databases (Donner, 2017; Yeung, 2019) but the metadata quality in these emerging databases is often lower than in the more established databases (Meho & Yang, 2007; Visser et al., 2021). By comparison, in Scopus and WoS, even though the *Data paper* document type is defined and used, papers bearing this label were introduced into these databases in different years, which contributes to inconsistent encoding of data papers. More importantly, we also find a stark gap in how publications in some of the data journals are encoded in these two databases. Through a more granular analysis using the case of *Scientific Data*, we find that the accuracy of this metadata element is significantly higher in WoS than in Scopus, which compensates for the less comprehensive coverage of data papers in the former data source.

Based on the results above, we argue that the inconsistent policies and implementation of the *Data paper* document type between popular scholarly databases pose a major issue for a more comprehensive understanding of the roles played by data papers in the research system. This is especially so given the fact that many of the data papers are published in mixed data journals, where data papers and research articles are published together and the document type is the only metadata field to easily distinguish these two genres. Future works should focus on how to more accurately label new and existing data papers to facilitate future quantitative studies.

**6 Conclusion**

In this research, we systematically investigated how data journals are indexed in four popular scholarly databases: the Web of Science, Scopus, Dimensions, and OpenAlex. We compiled a list of 18 exclusively data journals (i.e., journals that only or primarily publish data papers) by using the existing lists of data journals compiled by other researchers. Based on our list, we investigated how these journals and their publications are covered by the four databases, with particular attention to the document types used by the databases to index these data papers.

We find highly inconsistent coverage of data journals and papers as well as wide variation in how the *Data paper* document type is defined and assigned by these popular databases that are frequently used by researchers in quantitative studies of science. Emerging databases, such as Dimensions and OpenAlex, have much more comprehensive coverage of data journals and papers, even though WoS and Scopus still cover the most important journals and a significant amount of data papers. However, only the WoS and Scopus define a distinct *Data paper* document type. Moreover, this document type is assigned very inconsistently between the two databases. All these issues pose major barriers to comprehensive quantitative studies on this emerging academic genre, and more broadly, on the production and reuse of data in the research system. Hence, these issues should be addressed by databases and journals in the future to support a more open and data-centric research system.

As the next step of the project, we will use the list of exclusively data journals to investigate how the publishing and reusing of research data is connected to the discipline and gender of researchers. More importantly, we will also combine exclusively and mixed data journals to construct a large dataset of data papers, which will be critical for quantitative science studies on research data. Identification of data papers from mixed data journals would be another interesting research question to be solved in the future, particularly given the flaws in document type assignment that we observed in the major scholarly databases.